\documentclass[journal]{article}                                  

\usepackage{arxiv}
\usepackage{hyperref}
\usepackage{graphicx}          
\usepackage{amsmath} 
\usepackage{amssymb}  

\usepackage{amsthm}
\usepackage{color,soul}
\usepackage{flushend}
\usepackage{cite}
\usepackage{mfirstuc}
\usepackage{subfig}
\usepackage{mathtools}
\usepackage{xcolor}
\usepackage{array,multirow}
\usepackage{algorithm} 
\usepackage{algorithmic}  

\usepackage[linesnumbered,algo2e,ruled,vlined,norelsize]{algorithm2e} 
\usepackage[bb=libus]{mathalpha}
\usepackage{bbm}

\allowdisplaybreaks
\newcommand{\calW}{ \mathcal{W} }
\newcommand{\calA}{ \mathcal{A} }
\newcommand{\calN}{ \mathcal{N} }
\newcommand{\calD}{ \mathcal{D} }
\newcommand{\calT}{ \mathcal{T} }
\newcommand{\boldA}{ \mathbb{A} }
\newcommand{\soh}{\mathrm{SoH}_p}

\usepackage{tikz}
\usepackage{textcomp}
\usepackage{hyperref}
\usepackage{lipsum}

\usepackage{authblk}
\title{\LARGE \bf
Explainable Functional Relation Discovery for Battery State-of-Health Using Kolmogorov–Arnold Network
}

\author[1]{Sanchita Ghosh}
\author[1]{Tanushree Roy}
\affil[1]{Department of  Mechanical Engineering, Texas Tech University, Lubbock, TX 79409, US. Emails:~{\tt\small sancghos@ttu.edu, tanushree.roy@ttu.edu}.}

\begin{document}


\maketitle
\thispagestyle{empty}
\pagestyle{empty}


\begin{abstract}
Battery health management is heavily dependent on reliable State-of-Health (SoH) estimation to ensure battery safety with maximized energy utilization. Although online SoH estimation can effectively track battery degradation, it requires continuous battery data acquisition. In addition, model-based SoH estimation methods rely on accurate battery model knowledge, whereas data-driven approaches often suffer from limited interpretability. In contrast, analytical characterization of SoH will offer a direct and tractable handle on battery performance degradation, while also establishing a foundation for further analytical studies toward effective battery health management. Thus, in this work, we propose a  Kolmogorov–Arnold Network (KAN)-based data-driven pipeline to establish a functional relationship for SoH degradation using battery temperature data. Specifically, we learn long-term battery thermal dynamics and battery heat generation via learnable activation functions of our KAN model. We also propose a tailored loss function to incorporate physics-guided learning of the activation functions. We utilize this learned mapping to obtain an explicit functional relationship between SoH degradation and cycle number.  The proposed pipeline was validated using real-world data, yielding a closed-form analytical formula of SoH degradation with high accuracy.

\end{abstract}

\section{Introduction}

Lithium-ion batteries have become the preferred energy storage technology for applications ranging from consumer electronics and electric vehicles to grid-scale systems. However, repeated charge–discharge cycles produce irreversible power and capacity loss of the battery that can cause substantial degradation of performance.  Battery state-of-health (SoH) is a metric to assess the extent of battery degradation with aging and is often estimated as the usable-to-nominal capacity ratio, the present-to-initial discharge energy ratio, or the present-to-initial impedance ratio of the battery \cite{berecibar2016critical, preger2026capacity, zhou2018prognostics}. \cite{paradiso2026failure, sun2024safety} studied the increased risk of safety-critical incidents, including thermal runway, with battery aging and SoH drop. 
Particularly, \cite{zhou2021state} highlights the need for reliable SoH estimation to ensure safe and effective battery operations. 

Researchers have utilized measurable battery variables that are sensitive to variation in SoH to develop data-driven models for SoH estimation. For example,  \cite{zhang2023battery} developed a gated-recurrent-unit (GRU)-based SoH estimator that utilizes a gradually decreasing current in the constant voltage (CV) charging phase to extract features with double correlation analysis. Similarly, \cite{lin2012online} proposes an adaptive observer with online parameterization to track the long-term variation of the internal resistance for SoH estimation.  \cite{gao2024soh} adopts the Gaussian Process Regression (GPR) method with particle swarm optimization (PSO) for SoH estimation, while using partial charging voltage and current curves for health indicator extraction. \cite{gong2022state} utilizes constant current (CC) charging time-span, discharge duration, and incremental capacity curve to obtain ``health factors" and subsequently implements a long short-term memory network (LSTM)-PSO model to estimate SoH from them.  

Apart from the battery voltage and current characteristics, temperature characteristics have also been leveraged for SoH estimation. \cite{yang2022lithium} uses the change in battery surface temperature during the specific charge voltage interval to capture degradation in battery capacity. Similarly, \cite{tian2020state} adopts support vector regression to obtain a health indicator from differential temperature curves in a voltage range during CC charging for SoH estimation.   \cite{sun2024novel} proposed an attentional-feature-fusion integrated LSTM framework that uses differential temperature as the feature to estimate SoH with high accuracy under different operating temperatures.   Moreover, \cite{zhang2025cmmog} proposed a multi-task learning framework to combine health indicators from surface temperature and electrical characteristics to show the improved generalization capabilities of temperature-based health indicators. \cite{ahuja2026lithium} utilizes the battery surface temperature measurements to estimate a normalized heat generation term during CC charging and considers this normalized heat generation term as the health indicator for SoH estimation.  

In recent years, the Kolmogorov–Arnold Network (KAN) has been widely adopted in battery research due to its improved nonlinearity extraction capability and inherent interpretability (\cite{zhang2024lithium}). In particular, KAN has been used for state-of-charge estimation (\cite{sulaiman2024battery}),  core temperature prediction (\cite{mallick2025kan}), and diagnostics of battery thermal anomalies (\cite{ghosh2026kankoopman}).
Researchers have also explored KAN-based estimation of SoH. For instance, \cite{cheng2025soh} proposes a Convolutional Neural Network (CNN)-KAN model to estimate SoH from the capacity–power curve and battery temperature during CC-CV charging. Similarly, \cite{zhang2024lithium} develops a CNN-KAN-based SoH estimator using the charging time, current, and temperature from the CV charging phase that can adapt to varying charging current rates. \cite{he2025soh} utilizes partial discharge curve data for SoH estimation with a CNN-GRU-KAN model and shows that the model consistently performs well across various operating conditions.  \cite{chen2025estimation} develops a Transformer-integrated KAN model for SoH estimation from historical charge–discharge data and shows that the model can better capture the long-term temporal dependencies in the data. Similarly, \cite{jarraya2025soh} proposes an LSTM-KAN model where the LSTM module captures the long-term temporal dependencies and the KAN module better approximates the nonlinearity in the data.  Moreover, \cite{liu2025soh} combined the dynamic graph generator with KAN to estimate SoH for battery packs and shows that the incorporation of KAN improves the accuracy for capacity decline inflection point identification. These works highlight KAN's effectiveness in capturing the battery dynamics for estimation or prediction of battery states and its applicability for improved accuracy in SoH estimation. However, existing works combine KAN with Transformer or deep network-based modules such as CNN, GRU, or LSTM, and such integration loses the interpretability of KAN due to the complex architecture of the integrated modules. Thus, there is a research gap in utilizing the inherent interpretability of KAN to quantify battery degradation based on SoH estimation over the life-cycle of the battery.

From a practitioner’s perspective, while online SoH estimation from operational measurements remains important, developing a reliable, use-case-specific empirical model can provide a tractable quantification of battery degradation that is readily applicable to downstream analysis. Even though trends in SoH degradation can be inferred from online battery state estimation and subsequently modeled using parametric or non-parametric regression, such fitted functions often remain weakly grounded in the underlying battery degradation mechanisms.  While physics-based models can provide deeper understanding of underlying mechanism, they require comprehensive system knowledge. Thus, the first challenge in model-based SoH estimation is to properly identify the dynamics that accurately captures the SoH mechanism due to the complex coupling of system states arising from various internal reactions in the battery \cite{li2024soh}. Moreover, the accuracy of the model-based algorithms largely depends on reliable identification of the model parameters and requires separate parameterization for each battery. A recent work by \cite{ahuja2026lithium} leveraged the connection between progressive changes in heat generation and battery degradation to estimate SoH using a model-based online observer. However, the exact mapping between these terms remains implicit and can only be estimated based on continual measurement of battery temperature. In contrast, degradation quantification using data-driven techniques lacks interpretability and fails to provide analytical quantification of degradation trends. Therefore, there exists a research gap to obtain closed-form representations of SoH that are grounded in battery physics to preserve the interpretability of estimated battery degradation. 

To address these research gaps, the main contributions of this paper are as follows: 
\begin{enumerate}
    \item we provide a data-driven pipeline for utilizing battery temperature measurements to derive an analytical SoH characterization, 
    \item we quantify the increased heat generation in the battery in terms of the cycle number by learning the long-term thermal dynamics of the battery using Kolmogorov-Arnold Networks, 
    \item we propose a tailored loss function for effective quantification of heat generation from physics-guided learning of the activation functions, and
    \item we obtain an explicit functional relationship between SoH degradation and charging cycles by embedding the quantified heat generation.
\end{enumerate}

The rest of the paper is organized as follows: Section~\ref{preli} presents a brief overview of the KAN architecture and describes the impact of battery degradation on battery thermal dynamics. In Section~\ref{KAN soh}, we present our proposed strategy to utilize KAN-based learning of long-term variation in thermal dynamics, and the activation functions to capture the progressive increase in battery heat generation with degradation. Section~\ref{discovery_soh} discusses the implementation details of the KAN structure, followed by the results and findings. Finally, Section~\ref{conc} concludes our paper.

\section{Preliminaries}\label{preli}
In this section, we first present a brief overview of the KAN structure that leverages learnable activation functions to provide an interpretable data-driven model of a system.  Next, we describe how incremental heat generation with battery aging can be utilized to characterize the progressive degradation of SoH.

\subsection{Kolmogorov-Arnold network structure}
The Kolmogorov-Arnold network (KAN), proposed by \cite{liu2024kan}, is constructed as an $\calN$-layer deep network structure with $\calW_n$ width in each layer $n\in\calN_{kan},$ $\calN_{kan}=\{1, \hdots, \calN\}$, such that each node in a layer $n$ corresponding to a single input variable $x_{n,w},\, \forall w\in \calW_n,$ where $\calW_n=\{1,\hdots, W_n\},\, \forall n\in \calN_{kan}$. This architecture is built on the Kolmogorov–Arnold representation theorem, which states that any continuous multivariate function defined on a bounded domain can be represented as a finite sum of compositions of continuous univariate functions. For a nonlinear continuous function $f(z_1,\hdots, z_m)$, this theorem guarantees the existence of univariate functions $\calA_{p,q}, \overline{\calA}_{q}, p\in \{1, \hdots, m\}, q=\{1, \hdots, 2m+1\}$ such that 
\begin{align}\label{eq:funcn}
    f(z_1,\hdots, z_m) = \sum\limits_{q=1}^{2m+1} \overline{\calA}_{q}\left(\sum\limits_{p=1}^m \calA_{p,q}(z_p)\right).
\end{align}
Despite the guarantee of the existence of these univariate functions, their explicit construction remained challenging until the introduction of a learning-based deep architecture proposed by \cite{liu2024kan}. Specifically, KAN is trained to learn ``activation functions" $\calA_{n,\hat{w},w},\, \forall  w\in \calW_n, \hat{w}\in  \calW_{n+1}$ as a proxy for this univariate functions that map any input variables $x_{n,w}$ at a node $w$ to its output at node $\hat{w}$ in the consecutive layer $n+1$. These learnable activations with fixed additive operation at the nodes mark the primary difference between KAN and other standard neural networks, where the latter learn a linear weight combination of nodes with fixed activations. Mathematically, the output at each KAN node is given by
 \begin{align}
    x_{n+1,\hat{w}}=\sum_{w=1}^{W_n}\calA_{n,\hat{w},w} (x_{n,w}), \, \forall \hat{w}\in\calW_{n+1}.
\end{align}
 \noindent In a compact matrix form, it can be expressed as, 
\begin{equation} \label{KAN matrix}
    x_{n+1} = \underbrace{ \begin{pmatrix}
\calA_{n,1,1}(\cdot) & \dots & \calA_{n,1,W_{n}}(\cdot)\\
\vdots & \ddots & \vdots\\
\calA_{n,W_{n+1},1}(\cdot) & \dots & \calA_{n,W_{n+1},W_{n}}(\cdot)
    \end{pmatrix}}_{\boldA_{n}}x_{n},
\end{equation}
 where activation function matrix $\boldA_n$ denotes the mapping between the inputs $x_n=[x_{n,1}, \hdots, x_{n, W_n}]^T$ at layer $n$ and outputs $x_{n+1}=[x_{n+1,1}, \hdots, x_{n+1, W_{n+1}}]^T$ at the layer $n+1$. The input features to the KAN are often denoted as $x_0\in \mathbb{R}^{M_i}$, and the outputs of the KAN are denoted by $x_{N+1}\in \mathbb{R}^{M_0}$. In other words, the KAN architecture utilizes these activation functions $\boldA_n$ $\forall n \in \calN$ to capture the relationship between the $M_i$ input variables or features and  $M_{0}$ output or target variables as 
 \begin{align}
     x_{N+1}=KAN(x_0)=(\boldA_N\circ \boldA_{N-1}\circ \hdots \circ \boldA_1)x_0.
 \end{align}
 
 However, the above KAN architecture can be  simplified if the function $f$ in \eqref{eq:funcn} is known to be a sum of univariate functions such as 
 \begin{align}\label{eqn:simple_f}
  y=  f(z_1,\hdots, z_m) = \sum\limits_{p=1}^m \calA_{p}(z_p),
 \end{align}
which obviates the need for a $\calN$ deep architecture. A function defined as in \eqref{eqn:simple_f} can be represented in theory by a KAN architecture with a vector activation function $\boldA_1$ mapping $z = [z_1,\hdots, z_m]^T$ to $y$ as
\begin{align} \label{vector_activation}
    y=\boldA_{1}z = \begin{pmatrix}
\calA_{1,1,1}(\cdot) & \dots & \calA_{1,1,m}(\cdot)
    \end{pmatrix}z,
\end{align}
\subsection{SoH characterization using variation in heat generation}
Battery degradation with aging leads to a continual increase in the internal resistance of the battery. Power-based state-of-health metric, $SoH_P$, characterizes the degradation of the battery's capability to provide power due to this increase in the internal resistance. \cite{christen2023theory} showed that a small capacity fade due to the loss of an electrochemically active electrode surface is related to the loss of conductance ( or inversely related to resistance). This enables us to write the normalized inverse equivalent series resistance formula for $\soh$
\begin{align}
\soh\% = \frac{R_{\mathrm{BOL}}}{R} \times 100\%,
\label{eqn:soh_rs}
\end{align}
where $R$ is the equivalent series resistance of the battery and $R_{\mathrm{BOL}}$ represents the resistance at the beginning of the battery life. Now, such continual rise in the internal resistance also increases the heat generation in the battery over the cycles. Hence, battery thermal dynamics essentially captures the signature of battery degradation. 
For small-dimension single-cell batteries with a small Biot number, heat conduction inside the battery is much smaller than heat convection at the surface, so that the temperature gradient is negligible inside the battery \cite{mattia2025lithium}. Hence, the thermal dynamics of a cell, assuming a small Biot number, can be written as \cite{mattia2025lithium}
\begin{align}
\frac{dT}{dt} = - \frac{ hA\left(T - T_{\infty}\right) - I^2 R }{\rho c_p \nu},\, \forall t\in [0,\mathbb{T}_E],
\label{eqn:lumped}
\end{align}
where $T_s$ is the lumped battery temperature measured at the surface, $T_{\infty}$ is the ambient temperature, and $I$ is the current. The parameters $h$, $A$, $\rho$, $c_p$, and $\nu$ denote the convective heat transfer coefficient, cell surface area, density, specific heat capacity, and specific volume of the battery, respectively.  $\mathbb{T}_E$ here represents the time to reach the end-of-life (EOL) of the battery after the $E$-th cycle.

In \eqref{eqn:lumped}, the $I^2R$ term represents the battery heat generation that captures how the increase in internal resistance $R$ impacts the thermal behavior of the cell as it ages. To represent the aging behavior of the battery, the series resistance is assumed to change incrementally with battery cycling, i.e., remains unchanged over any cycle $k$. Therefore, we can assume here that the battery heat generation term $ \zeta (k) = I^2R(k)$  remains constant during the $k$-th constant-current (CC) charging cycle, where $R(k)$ is the internal resistance at the $k$-th cycle. 
Then, \eqref{eqn:soh_rs} can be equivalently written in terms of $R(k)$ for any cycle $k$ as: 
\begin{align}\label{eqn:sohR_k}
    \soh(k)\% = \frac{\zeta_{\mathrm{B}}}{\zeta(k)} \times 100\%,
\end{align}
Here, $\zeta_{\mathrm{B}} = I^2R_{\mathrm{BOL}}$ is the heat generation at the beginning of battery life. Next, we present a hybrid lumped battery thermal model  that captures the long-term evolution of temperature as well as the progressive heat generation over the battery cycles.  

\subsection{Long-horizon battery thermal model for KAN learning}
Learning the data-driven representation of the lumped thermal model requires identifying the discretized model for \eqref{eqn:lumped} in terms of normalized variables. Let $\{\calT_k\}_{k=1}^{E}$ be a non-uniform distribution of points within $[0,\mathbb{T}_E]$, where $0< \calT_0 < \calT_1 < \cdots < \calT_E < \mathbb{T}_E$, such that each interval $[\calT_{k-1},\calT_{k})$ denote the $k$-th battery charging cycling time interval, $\forall k\in\{1,\hdots, E\}$. We will represent three sets of time samples from each of these cycling intervals
\begin{align}
    I_{\text{train}} &= \{t_k: {t}_k \in [\calT_{k-1},\calT_{k}), \forall k\in \{1, \hdots, E\} \},\\
    I_{\text{validation}} &= \{\overline{t}_k: \overline{t}_k \in [\calT_{k-1},\calT_{k}), \forall k\in \{1, \hdots, E\} \},\\
    I_{\text{test}} &= \{\widetilde{t}_k: \widetilde{t}_k \in [\calT_{k-1},\calT_{k}), \forall k\in \{1, \hdots, E\} \},
\end{align}
that represent the data samples for training, validation, and testing, respectively.
To analyze the training process, we will write \eqref{eqn:lumped} in terms of min-max normalization, where the normalized battery temperature is given by 
\begin{align} \label{eqn:normal_T}
    \overline{T}(t) = \tfrac{T(t)- T_{min}}{T_{max}- T_{min}},\,\forall t\in [0,\mathbb{T}_E],
    \end{align}
    where 
    \begin{align}
    &T_{max} = \max_{t\in I_{\text{train}}}[T(t)] \text{ and }T_{min} = \min_{t\in I_{\text{train}}} [T(t)],
    \end{align}
    are the maximum and minimum temperatures over the training dataset. Since $T_{max},T_{min}$ are constants, we can write $\tfrac{d \overline{T}}{dt} = \frac{1}{\Delta_T}\frac{d {T}}{dt}$, where, $\Delta_T = T_{max}- T_{min}$. Then, using the definition of normalized temperature from \eqref{eqn:normal_T} in \eqref{eqn:lumped}, we can write the $\overline{T}(t)$ dynamics $\forall t\in [0,\mathbb{T}_E]$ as:
\begin{align}
\frac{d \overline{T}}{dt} =- 
\frac{hA}{\rho c_p \nu}\overline{T} +\frac{hA}{\rho c_p \nu \Delta_T}\left( T_{\infty}- T_{min} \right) + \frac{I^2 R}{\rho c_p \nu \Delta_T}.
\label{eqn:lumped_sub}
\end{align}
In discrete form, \eqref{eqn:lumped_sub} can then be represented as
\begin{align}
    \overline{T}(i+1) = \gamma\overline{T}(i) + \xi + \frac{\tau}{\rho c_p \nu \Delta_T} I^2 R(i),\label{eqn:discrete1}
\end{align}
where $T(i+1)$ represents the subsequent temperature measurement at a time $\tau$ seconds later than the measured temperature $T(i)$ and $\tau$ denotes the sampling time. Here, constants $\gamma:=\left(1-\tfrac{hA\tau}{\rho c_p \nu }\right)$ and $\xi := \tfrac{hA\tau}{\rho c_p \nu \Delta_T}\left( T_{\infty}- T_{min} \right)$. 
We will use the lumped thermal model \eqref{eqn:discrete1}  to derive the long-term evolution of battery temperature over a fixed time horizon of $H\ll\min_k|\calT_{k}-\calT_{k-1}|$. Let us define
\begin{align} \label{eqn:Dk}
    \calD_k := \frac{\tau}{\rho c_p \nu \Delta_T}I^2R(i_k),\, i_k\in [\calT_k,\calT_{k+1}),
\end{align}
such that $\calD_k$ corresponds to the heat generation during the $k$-th CC charging cycle of the battery. 
Then, for any $k$-th CC charging cycle and $i_k, i_{k}+H \in [\calT_k,\calT_{k+1})$, the long-horizon hybrid model can be written in terms of $\calD_k$ as
\begin{align}
   \overline{T}(i_k+H)& = \gamma^N \overline{T}(i_k) +  \xi \sum\limits_{j=0}^{N-1} \gamma^j + \calD_k  \sum\limits_{j=0}^{N-1} \gamma^j,  \label{eqn:discrete}
\end{align}
where $H=N\tau$. Additionally, using \eqref{eqn:Dk} and the definition of $\zeta_k$, \eqref{eqn:sohR_k} can be equivalently written as:
\begin{align}\label{eqn:sohD}
    \soh(k)\% = \frac{\calD_{{B}}}{\calD_k} \times 100\%,
\end{align}
  where $\calD_{{B}}$ corresponds to the heat generation at the beginning of the battery life. 
We will first use this long-term prediction formula for normalized temperature $\overline{T}(i_k+H)$ to construct the architecture of the KAN model, and utilize the learned mapping to quantify the increased heat generation in the battery to estimate $\soh$ as \eqref{eqn:sohD}.
\begin{figure}[t!]
    \centering
    \includegraphics[width=0.75\linewidth]{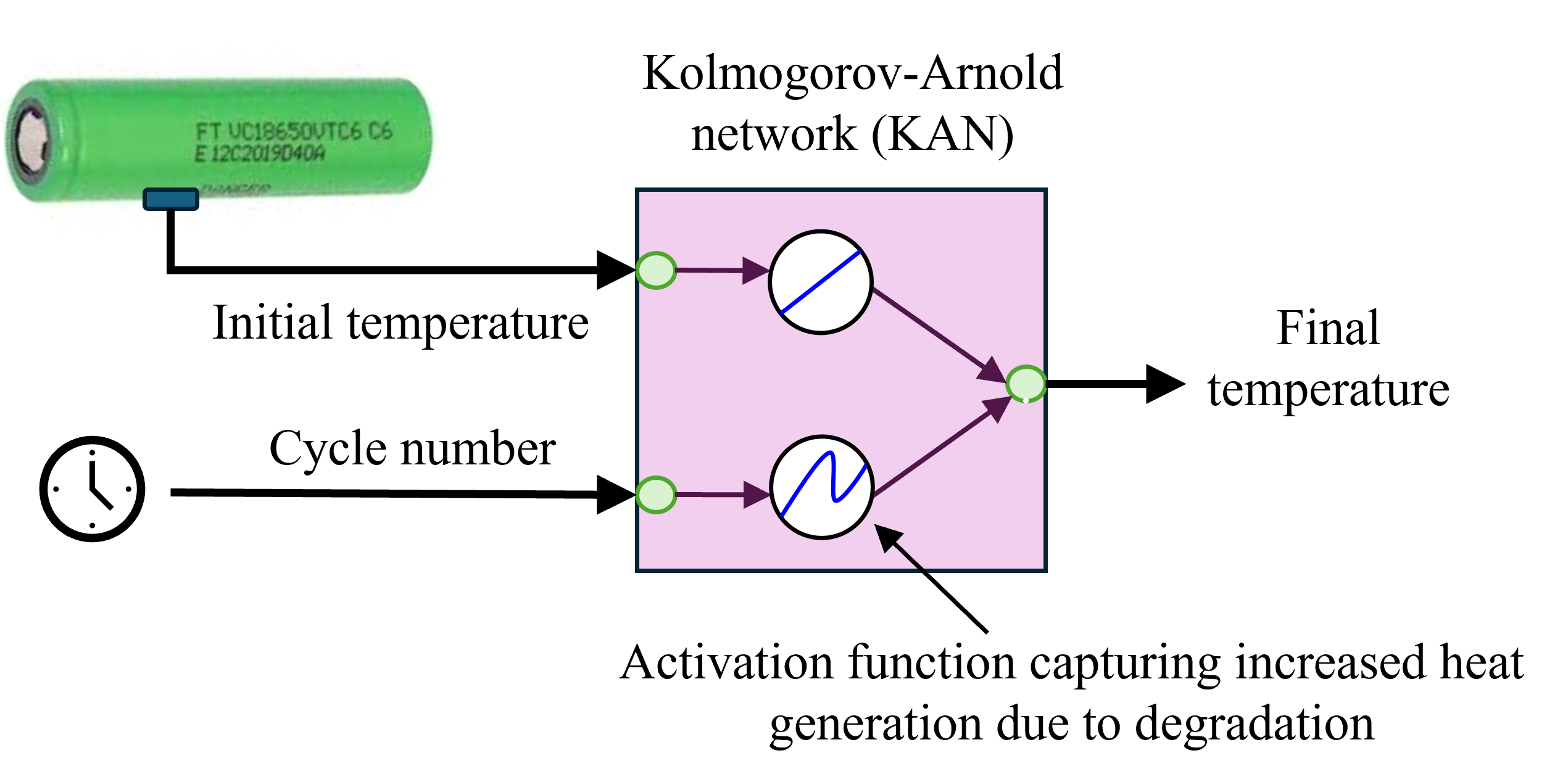}
    \caption{Overview of the data-driven pipeline used for discovering the functional relationship between battery degradation and battery cycle number.}
    \label{fig:overview}
\end{figure}

 \section{Kolmogorov-Arnold Network For SoH Estimation} \label{KAN soh}
In this section, we first present our proposed structure of an interpretable KAN model to subsequently isolate the increased heat generation over charging cycles due to battery degradation. Next, we discuss the significance of this KAN model in comparison to the state-of-the-art data-driven methods.
\subsection{Discrete activation functions for SoH estimation}
 Our goal in this paper is to leverage the inherent interpretability of the KAN structure to establish an explicit functional formula for $\soh$ in terms of the cycle number $k$. Therefore,  following \eqref{vector_activation}, we develop a 1-layer KAN model of width $[\![2, 1 ]\!]$ that has an activation function vector $\boldA = \{\calA_{1}, \,\, \calA_{2}\}$. This KAN model maps the battery thermal dynamics as an addition of activation functions $\calA_{1} $ and $\calA_{2} $ such that  \eqref{eqn:discrete} can be written as
\begin{align}
    &\overline{T}(i_k+H) = \calA_{1} (\overline{T}(i_k))+\calA_{2}(\overline{k}),  \label{kan learn}\\
&\calA_{1}(\overline{T}(i_k)) =  \gamma^N \overline{T}(i_k) +  \xi \sum\limits_{j=0}^{N-1} \gamma^j,\label{eqn:A1}\\
    &\calA_{2}(k) = {\calD_k} \sum\limits_{j=0}^{N-1} \gamma^j.\label{eqn:A2}
\end{align}
Here, $\overline{k} = \frac{k}{E}$ is the scaled cycle number, and the identified activation function $\calA_{2}(k) $ corresponds to the heat generation during the $k$-th CC charging cycle of the battery.
Hence, $\soh$ in \eqref{eqn:sohD} can be written in terms of the functions $\calA_{2}(k)$
\begin{align}
    \soh(k)\% =& \frac{\calA_{2}(0)}{\calA_{2}(k)} \times 100\%. \label{eqn:dirA}
\end{align}
Fig~\ref{fig:overview} illustrates the proposed data-driven pipeline for obtaining an explicit functional relation between SoH degradation and battery cycling number, while utilizing the learnable activation function of the KAN structure to capture the increased heat generation with degradation. 

Although such 1-layer KAN structure preserves the interpretability of the learned mapping of the long-horizon thermal dynamics \eqref{eqn:discrete} as \eqref{kan learn}, it requires task-specific training strategies to ensure physics-compliant learning of the activation functions as well as to attain high predictive accuracy.  Thus, during training, we minimize a tailored loss function \eqref{estimation loss} that penalizes the prediction error and the activation functions. 
\begin{align} 
\ell_{total} &= \lambda_{pred} \ell_{pred} + \lambda_{hess} \ell_{hess} + \lambda_{\soh} \ell_{\soh}, \label{estimation loss} \\
\ell_{pred} & = \frac{1}{M} \sum \limits_{j= 1}^M {\left( \overline{T}(j) - \widehat{\overline{T}}(j) \right)^2}, \label{mse_pred} \\
\ell_{hess} & = \frac{1}{M} \sum \limits_{j= 1}^M\bigg\lvert   \frac{\calA_{1}(\overline{T}(j))}{\overline{T}(j)}  \bigg\rvert, \label{hess_loss} \\
\ell_{\soh} & = \frac{1}{M} \sum \limits_{j= 1}^M\bigg\lvert   \frac{SoH_{p,j}}{100} - \frac{\calA_{2}(0)}{\calA_{2}(\overline{k}_j)} \bigg\rvert. \label{soh_loss} 
\end{align}
Here, $\ell_{pred}$ captures the mean-squared  error in temperature prediction, $\ell_{hess}$ includes the mean second-order derivative of the activation function $\calA_{1}$, and $\ell_{\soh}$ represents the mean-absolute error in $\soh$ estimation using $\calA_{2}$. $\lambda_{pred}$, $\lambda_{hess}$, and $\lambda_{\soh}$ are the learnable weights for the prediction, hessian, and $\soh$ losses, respectively. The hessian loss  $\ell_{hess}$ promotes the expected linearity in $\calA_{1}$ from $\gamma^N$ \eqref{eqn:A1}, and $\soh$ loss $\ell_{\soh}$ ensures that $\calA_{2}$ captures the degradation-induced incremental heat generation in the battery. Thus, minimizing $\ell_{total}$ improves overall predictive accuracy as well as enforces physics-based constraints on activation functions of the 1-layer KAN model.

 \subsection{Relevance of KAN as an interpretable model}
Model interpretability is the measure of the understanding of the model's reasoning behind its prediction for a given input. Thus, a model is considered interpretable if we can comprehend the mapping from the model's input to its prediction. The KAN architecture is inherently interpretable, since it learns the functional relation between the input features and the output data as a composition of activation functions. These activation functions, learned as a linear combination of silu-base functions and $k$-order B-splines over $G$ grid points, can be isolated to identify the contribution of each feature to the final output or prediction. 

Specifically, for the KAN structure of width $[\![2, 1 ]\!]$ chosen in our study, we learn two activation functions, each corresponding to one input feature.  {The first function $\calA_{1}$ maps the initial normalized temperature $\overline{T}(i_k)$   to the next step normalized temperature $\overline{T}(i_k+ H)$, and $\calA_{2}$ captures the functional mapping from the normalized cycle number $\overline{k}$ to the normalized temperature $\overline{T}(i_k+H)$.} Moreover, the KAN training also incorporates regularization to avoid overfitting the B-splines and reduce model complexity. While this constraint in training encourages learning simpler functional relations, the piecewise nature of the B-splines may still make them intractable. Thus, we identify functions that closely approximate these learned splines using a broad range of functional dictionaries containing polynomial, exponential, trigonometric, logarithmic, and similar standard basis functions. Although these advantages highlight the relevance of KAN as an interpretable model, we now present the reasons for selecting it as our preferred approach over other comparable data-driven methods.

\textbf{I. Parametric Regression:} These regression methods can be utilized to explicitly learn the functional relations between $\soh$ and cycle number $k$ (\cite{chen2025residual}). However, such mapping lacks any insight regarding battery degradation. In contrast, with KAN, we can capture the impact of battery degradation through the increased heat generation in the battery.

\textbf{II. Neural Networks (NN):}
NN-based models learn the data pattern by successively applying nonlinear transformations to linear combinations of weights such that the underlying data structures get embedded into relatively larger representations. This model complexity decreases model interpretability, and the nonlinear transformation at each node impedes the isolation of the individual contribution of input features, even for simpler models. Thus, traditional NN models such as MLP, LSTM, and RNN will not be suitable for our objective.

\textbf{III. Tree-based Learners:} Tree-based models such as random forest, XGBoost, light gradient boosting machine utilize rule-based splits to capture the underlying data structures (\cite{yang2022prediction}). However, it is difficult to trace each split to understand the mapping of a tree-based model's input to output and often requires post-analysis tools (\cite{xiao2024state}. Thus, these models remain less interpretable and fail to provide compact representations for the learned functions. Moreover, they also lack the ability to isolate the feature-specific impact on the model output. 

\textbf{IV. Gaussian Process Regression (GPR):}
 GPR method adopts a kernel-based learning strategy to identify the probability distribution over all possible functions to fit the training data.  GPR can provide insight about the underlying patterns in data; however it fails to provide closed-form representations of the learned functions (\cite{jia2020soh}).

\textbf{V. Generalized Additive Models (GAMs):} GAMs, like KAN, learn the patterns in data using compositions of univariate functions. However, GAMs only consider the spline function as the basis, while the KAN  utilizes a sum of silu and spline functions. Incorporating the silu basis function helps the KAN  to identify the dominant trend in the input contribution, and the KAN framework also supports improved scalability both in terms of grid size and depth of the network compared to GAMs (\cite{zhang2026kan}).

\section{Discovery of Explicit Functional Formula for SoH}\label{discovery_soh}
In this section, we present the details of estimating the functional formula for $\soh$ with cycle number using real-world battery datasets.

 \subsection{Description of KAN training}
We validate our proposed KAN-based data-driven pipeline for two real-world datasets. First, we consider three battery data: VAH09, VAH137, and VAH17 from \cite{bills2023battery} where these three cells exhibit very different degradation trends. \cite{bills2023battery} provides battery aging data under electric vertical takeoff and landing (EVTOL) current profiles for Sony-Murata 18650 VTC-6 battery cells. These cells have a rated capacity of $3000mAh$ at a nominal voltage of $3.6V$. VAH09 and VAH17 batteries are cycled with a baseline current profile at $20^\circ C$ and $23^\circ C$, respectively. VAH13 cell is cycled with reduced cruise time at $23^\circ C$.  Next, we consider cell3 from the \cite{howey2017oxford} dataset to evaluate our proposed method for degradation under dynamic current discharging of the battery. This is a Kokam (SLPB533459H4) pouch cell battery with 740 mAh rated capacity and $3.6V$ nominal voltage. The cell is charged with 2CC current and discharged with the Artemis urban driving profile \cite{andre2006real}. 
For all battery datasets, we follow the steps below to generate our training, validation, and testing datasets.

\textbf{Training Dataset:} To capture the long-term evolution of normalized temperature, we first uniformly subsample the normalized surface temperature measurements from the training sample interval $I_{\text{train}}$ over a long-horizon $H$ within the CC charging phase of each cycle. Then we use the normalized temperature at the $i^{th}$ and the scaled cycle number $\overline{k}$ as input features to predict the normalized temperature at the $(i+H)^{th}$ of this $k^{th}$ cycle. Thus, we create the input dataset $D_{x}$ and the corresponding target dataset $D_{y}$ for KAN training as below.
\begin{align}
    &D_{x} = \{x (0), \quad\quad \, \,\, \,x(1), \quad\quad \,\,\,\cdots, \,\,  x (E)\}, \label{dz dataset} \\
    &D_{y} =  \{\overline{T}(i_0+H), \,\, \overline{T}(i_1+H),\,\,\,\cdots, \,\overline{T}(i_E+ H)\}, \label{dy dataset}\\
    &\text{where,}\,\, x (k) = [\overline{T}(i_k) \,\quad \overline{k}]^T. \nonumber 
\end{align}

\begin{figure}[h!]
    \centering
    \includegraphics[width=0.9\linewidth]{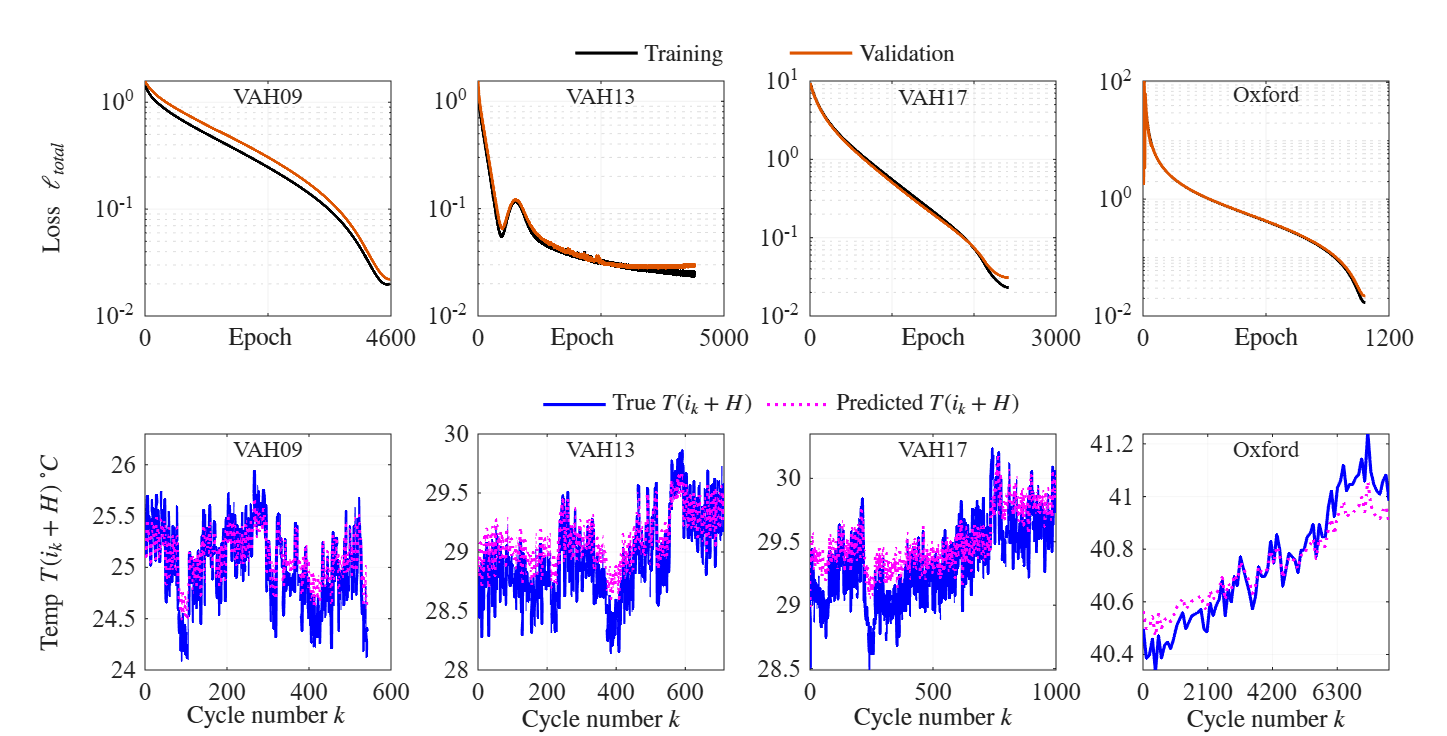}
    \caption{The top plots show the converging training and validation losses during the training of four KAN models corresponding to VAH09, VAH13,  VAH17, and Oxford batteries (from left to right). The bottom plots show the true and predicted long-term prediction of lumped battery temperature, measured at the surface for each battery.}
    \label{fig:tempTR_8sub}
\end{figure}
\textbf{Validation Dataset:} We similarly create the validation datasets $\overline{D}_{x}$ and $\overline{D}_{y}$ by sub-sampling the normalized surface temperature measurements from the validation sample interval $I_{\text{validation}}$ over a long-horizon $H$.
\begin{align}
    &\overline{D}_{x} = \{\overline{x} (0),  \qquad\,\, \,\,\cdots, \,  \overline{x} (E)\}, \label{dz vld dataset} \\
    &\overline{D}_{y} =  \{\overline{T}(\overline{i}_0+H), \,\cdots, \,\overline{T}(\overline{i}_E+ H)\}, \label{dy vld dataset}\\
    &\text{where,}\,\, \overline{x} (k) = [\overline{T}(\overline{i}_k) \,\quad \overline{k}]^T. \nonumber 
\end{align}

\textbf{Testing Dataset:} We sub-sample the normalized surface temperature measurements from the testing sample interval $I_{\text{test}}$ over a long-horizon $H$ of CC battery charging to create the datasets $\widetilde{D}_{x}$ and $\widetilde{D}_{y}$ for testing of the learned KAN model.
\begin{align}
    &\widetilde{D}_{x} = \{\widetilde{x} (0),  \qquad\, \,\,\cdots, \,  \widetilde{x} (E)\}, \label{dz tst dataset} \\
    &\widetilde{D}_{y} =  \{\overline{T}(\widetilde{i}_0+H), \,\cdots, \,\overline{T}(\widetilde{i}_E+ H)\}, \label{dy tst dataset}\\
    &\text{where,}\,\, \widetilde{x} (k) = [\overline{T}(\widetilde{i}_k) \,\quad \overline{k}]^T. \nonumber 
\end{align}

\renewcommand{\arraystretch}{1.3} 
\begin{table}[h!]
    \centering
    \caption{Summary of hyperparameters and training performance metrics for KAN models.}
    \begin{tabular}{|c|c|c|c|c|}
    \hline
    Parameter & VAH09 & VAH13 & VAH17 & Oxford \\
    \hline
      Grid   & 8 & 8 & 5 & 8\\
      \hline
       Spline-order $\kappa$  & 6 & 7 & 6 & 8 \\
       \hline
       Learning rate  & 8e-4 & 5e-4 & 5e-4 & 5e-4\\
       \hline
       $\lambda_{pred}$ & 1 & 1 & 1 & 1\\
       \hline
       $\lambda_{hess}$ & 0.6 & 0.7 & 0.4 & 0.6 \\
       \hline
       $\lambda_{\soh}$ & 1 & 0.9 & 0.8 & 0.9 \\
       \hline
        Epoch & 4582 & 4868 & 2512 & 1136 \\
        \hline
        Training data size & 542 & 712 & 998 & 800 \\
        \hline
        Training Loss & 0.019 & 0.024 & 0.023 & 0.017 \\
        \hline
        Validation Loss & 0.022 & 0.026 & 0.031 & 0.021\\
        \hline
        Test MAE & 0.132$^\circ C$ & 0.168$^\circ C$ & 0.179$^\circ C$ & 0.103$^\circ C$\\
        \hline
    \end{tabular}
    \label{tab:training parameters}
\end{table}
We train a dedicated 1-layer KAN model of width $[\![2, 1 ]\!]$ for each of the four batteries using their corresponding training datasets ${D}_{x}$, ${D}_{y}$ along with validation datasets $\overline{D}_{x}$ and $\overline{D}_{y}$.  We represent the activation functions $\calA_{1}$ and $\calA_{2}$ with a sum of SiLU functions and a linear combination of  B-splines. To effectively learn these functions, we minimize the loss function $\ell_{total}$ \eqref{estimation loss} using the `Adam' optimizer. All models are trained using full-batch gradient descent, without employing mini-batch training. Additionally, we conducted an exhaustive grid search for each KAN model to determine the suitable hyperparameter combination based on the validation loss.  Table~\ref{tab:training parameters} lists the selected hyperparameters, together with training and validation losses and mean absolute error (MAE) for temperature prediction in the testing datasets.

The top plots of Fig~\ref{fig:tempTR_8sub} show the convergence of training and validation over the training steps for each KAN model. We test these KAN models using the test datasets $\widetilde{D}_{x}$ obtained from these batteries, and the bottom plots of Fig~\ref{fig:tempTR_8sub} compare the true and predicted battery temperatures. The left three plots in the bottom highlight that the KAN models exhibit an over-prediction trend with MAE of 0.132$^\circ C$, 0.168$^\circ C$, and 0.179$^\circ C$, respectively, for the VAH09, VAH13, and VAH17 batteries that are cycled with a CC profile. However, the bottom right plot shows that the KAN model over-predicts during early cycles and under-predicts for the aged condition for the Oxford battery, and achieves an MAE of 0.103$^\circ C$.  Most importantly, Fig~\ref{fig:tempTR_8sub}  indicates that the KAN models effectively learned the progressive increase in temperature with aging, due to the battery degradation.

\subsection{KAN activation function-based SoH estimation } 
From \eqref{eqn:discrete} and \eqref{eqn:A2}, we observe that the activation function  $\calA_{2}(k)$ maps the heat generation term ${\calD_k}$. Thus, we utilize $\calA_{2}(k)$ from our KAN models to estimate $\soh$ as \eqref{eqn:dirA}, where $\calA_{2}(k)$ is learned as a composition of SiLU and B-spline functions. We directly evaluate the learned activation function $\calA_{2}(k)$ in the range $k\in [0, \cdots, E]$. We compare this B-spline-based estimated $\soh$  with a baseline $\soh$ defined as \eqref{eqn:soh_rs}.  We utilize the voltage and current measurement data for each battery to obtain the resistance $R$ for each cycle with the IR drop method as $R = \tfrac{\Delta V}{\Delta I}$. Thus, we use the instantaneous voltage jump corresponding to one current step at the beginning of the CC charging phase for each cycle to calculate the battery resistance, and subsequently use these resistance $R$  values to calculate the baseline $\soh$ using \eqref{eqn:soh_rs}. Next, we evaluate the proposed $\soh$ estimation method for different batteries:  18650 VTC-6 and Kokam pouch cells, and under diverse cycling conditions.

\textbf{Performance under constant current cycling:}
\begin{figure}[h!]
    \centering
    \includegraphics[width=0.9\linewidth]{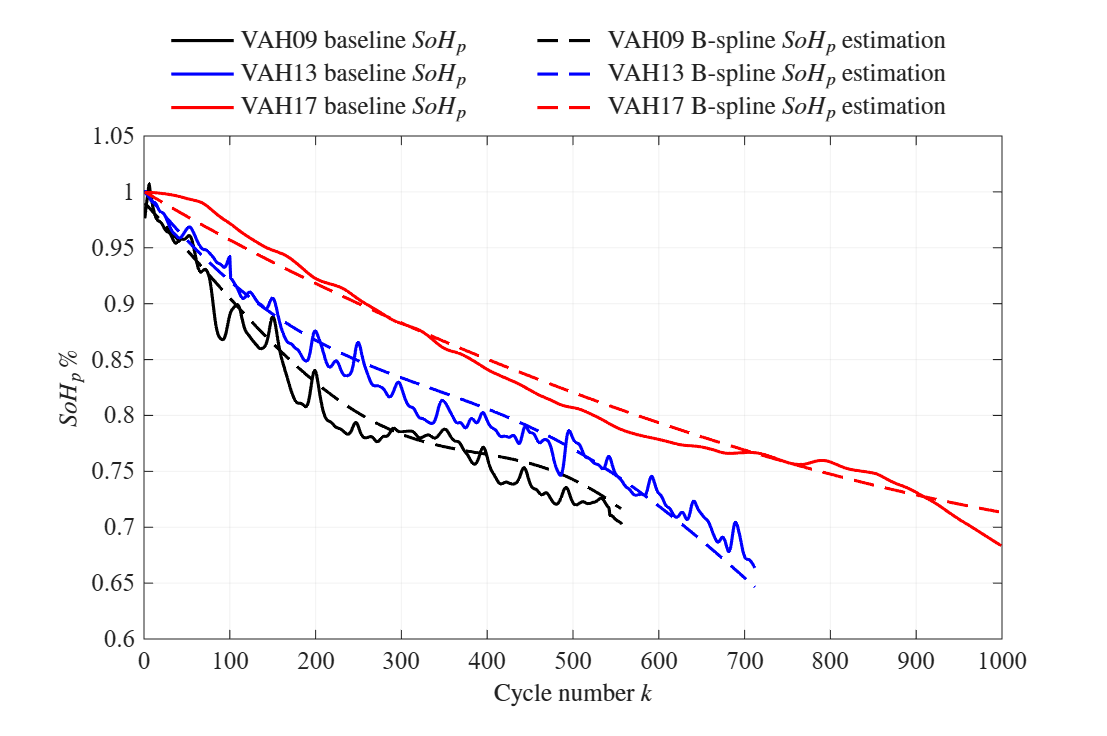}
    \caption{Plot shows the baseline and the estimated $\soh$ for three batteries cycled under different CC profiles and ambient temperatures. The true $\soh$ is calculated using the drop in resistance. The $\soh$ estimations are calculated from the B-spline representation of $\calA_{2}(k)$. }
    \label{fig:sohpred}
\end{figure}

We first evaluate the proposed $\soh$ estimation pipeline for the VAH09, VAH13, and VAH17 batteries from the \cite{bills2023battery} dataset. These  three batteries are cycled under different CC cycling profiles and ambient temperatures, and consequently, they exhibit distinct degradation trends. Fig~\ref{fig:sohpred} shows the baseline and estimated $\soh$ graphs, where the baseline $\soh$ graphs are plotted with black, blue, and red lines, respectively, for the VAH09, VAH13, and VAH17 batteries, and their corresponding B-spline $\soh$ estimates are plotted with the same color dashed lines. Fig~\ref{fig:sohpred} highlights that the KAN-based $\soh$ estimation closely follows the baseline for each battery. VAH09 battery exhibits the shortest life span with EOL  at the 557$^{th}$ cycle and $70.2\%$ $\soh$. For this battery,  B-spline $\soh$ estimation achieves an MAE of $1.35\%$, a maximum absolute error (MaxAE)  of $4.24\%$,  and $1.3\%$ overestimation error at EOL.  VAH13 battery reaches EOL at the 712$^{th}$ cycle with $66.4\%$ $\soh$, and the  B-spline $\soh$ estimation exhibits $1.14\%$ MAE,  $4.18\%$ MaxAE, $1.71\%$ under-estimation error at EOL. Additionally, the B-spline $\soh$ estimation for VAH17  achieves an MAE of $0.94\%$, a MaxAE of $3.0\%$, and  $3.0\%$ overestimation error at EOL, where the battery reaches EOL at 998$^{th}$ cycle with $68.4\%$ $\soh$. Overall, Fig~\ref{fig:sohpred} indicates that the KAN models effectively capture the degradation-driven progressive heat generation in each battery with the activation function $\calA_{2}(k)$, leading to an accurate $\soh$ estimation.

\renewcommand{\arraystretch}{1.5} 
  \begin{table*}[b!]
        \centering
         \caption{Compares  $\soh$ estimation performance with selected functional representations of $\calA_2$ }
        \begin{tabular}{|c|c|c|c|c|c|c|c|c|}
        \hline
          Battery   & Metric & B-splines & Square & Cubic & Quartic   & Quintic & Exp & Trig \\
          \hline \hline
          
          \multirow{5}{*}{VAH09}   & MAE  & 1.35 & 1.36 & 1.33 & 1.37 & 1.38 &  1.43 & 1.38 \\
          \cline{2-9}
              & MaxAE  & 4.24 & 4.32  & 4.21 & 4.26 & 4.36 & 4.43 & 4.26 \\
              \cline{2-9}
               & RMSE  & 1.65 & 1.68 & 1.62 & 1.68 & 1.70 & 1.76 & 1.69 \\
               \cline{2-9}
                & EOL-E  & 1.30 & 1.34 & 1.33 & 1.32 & 1.34 & 1.33 & 1.42 \\
                \cline{2-9}
                & $R^2$ score  & - & 0.99982 & 0.99983 & 0.99981 & 0.99981 & 0.99980 & 0.99982 \\
                \hline \hline
          
          \multirow{5}{*}{VAH13}   & MAE  & 1.14 & 1.28  & 1.26 & 1.30 & 1.36 & 1.28 & 1.21 \\
          \cline{2-9}
              & MaxAE  & 4.18 & 4.26  & 4.23 & 4.27 & 4.27 & 4.28 & 4.22 \\
              \cline{2-9}
               & RMSE  & 1.47 & 1.64 & 1.59 & 1.68 & 1.75 & 1.63 & 1.52 \\
               \cline{2-9}
                & EOL-E  & 1.71 & 1.70 & 1.74 & 1.76 & 1.77 & 1.79 & 1.83 \\
                \cline{2-9}
                & $R^2$ score  & - & 0.98674 & 0.98697 & 0.98646 & 0.98528 & 0.98620 & 0.98717 \\
                \hline \hline
          
          \multirow{5}{*}{VAH17}   & MAE  & 0.94  & 1.02 & 0.98 & 1.0 & 1.02 & 0.97 & 1.03 \\
          \cline{2-9}
              & MaxAE  & 3.0 & 3.01 & 2.98 & 2.99 & 3.01 & 2.99 & 3.02 \\
              \cline{2-9}
               & RMSE  & 1.11 & 1.38 & 1.31 & 1.36 & 1.37 & 1.28 & 1.41 \\
               \cline{2-9}
                & EOL-E  & 3.0  & 3.01 & 2.98 & 2.99 & 3.01 & 2.99 & 3.02 \\
                \cline{2-9}
                & $R^2$ score  & - & 0.99876 & 0.99923  & 0.99884 & 0.99871 & 0.99901 & 0.99846 \\
                \hline \hline
          
          \multirow{5}{*}{Oxford}   & MAE  & 0.7 & 0.81 & 0.78 & 0.78 & 0.80 & 0.74 & 0.83 \\
          \cline{2-9}
              & MaxAE  & 2.43 & 2.48 & 2.42 & 2.46 & 2.47 & 2.48 & 2.51 \\
              \cline{2-9}
               & RMSE  & 0.91 & 1.12 & 1.03 & 1.02 & 1.08 &  0.98 & 1.17 \\
               \cline{2-9}
                & EOL-E  & 0.27 & 0.26 & 0.29 & 0.28 & 0.27 & 0.30 & 0.28 \\
                \cline{2-9}
                & $R^2$ score  & - & 0.99351 & 0.99437 & 0.99435 & 0.99434 & 0.99415 & 0.99321 \\
                \hline
        \end{tabular}
       
        \label{tab:all_bat soh result}
    \end{table*}

    \textbf{Performance under dynamic current cycling:}
    \begin{figure}[h!]
    \centering
    \includegraphics[width=0.9\linewidth]{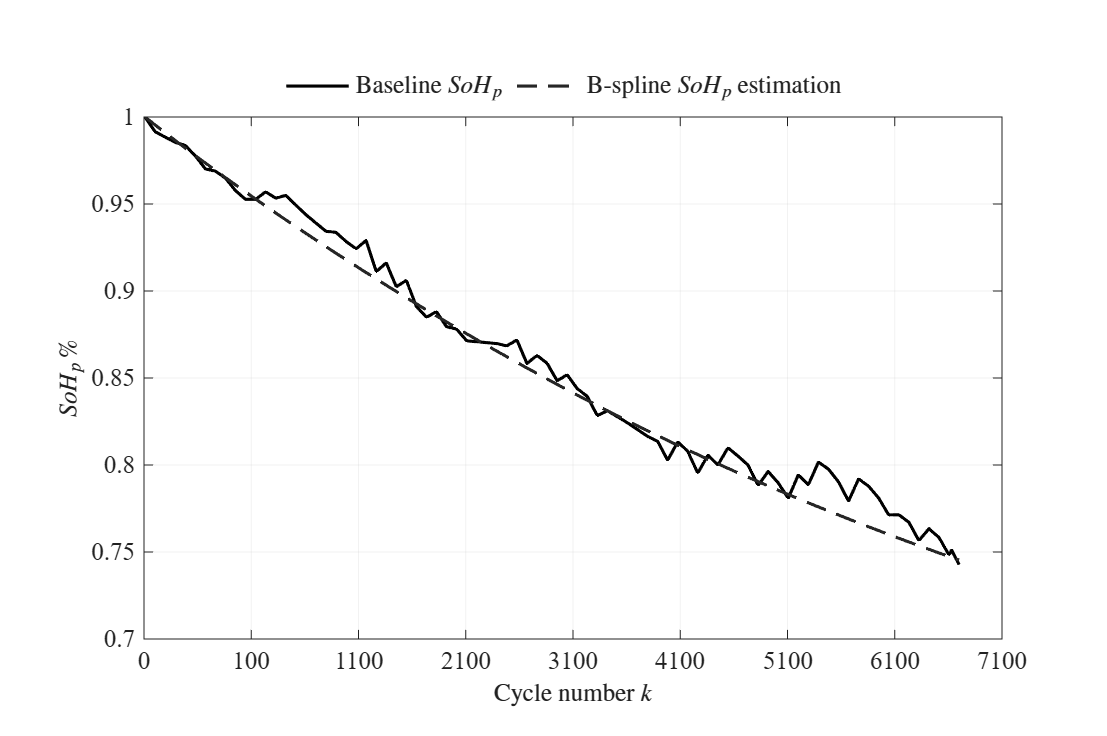}
    \caption{Plot shows the baseline and the estimated $\soh$ for the Oxford battery cycled under the Artemis urban driving current profile. The true $\soh$ is calculated using the drop in resistance. The $\soh$ estimations are calculated from the B-spline representation of $\calA_{2}(k)$. }
    \label{fig:sohpredOxford}
\end{figure}
The battery cell from the Oxford dataset is cycled under the Artemis urban driving profile, with reference measurements for $\soh$ calculation at every 100$^{th}$ cycle. Fig~\ref{fig:sohpredOxford} captures the baseline and estimated $\soh$ graphs for this cell, plotted with black solid and dashed lines, respectively. The cell reaches EOL at the 8100$^{th}$ cycle with $74.3\%$ $\soh$.  Fig~\ref{fig:sohpredOxford} shows that the B-spline $\soh$ estimation nearly matches the baseline $\soh$, and yields an MAE of $0.7\%$, a MaxAE of $2.43\%$, and  $.27\%$ over-estimation error at EOL.

These results demonstrate that our proposed KAN-based pipeline for $\soh$ estimation exhibits consistent performance for different batteries and cycling conditions, including constant and dynamic current profiles and varied ambient temperatures.
 \subsection{Closed-form representation of SoH degradation} 
 The B-spline representation of $\soh$ using $\calA_2$ provides actionable insights into the incremental heat generation that are grounded in battery physics.
 Our results show that they also yield reliable estimation accuracy. However, it is difficult to comprehend or conduct further mathematical analysis with these piecewise-learned B-splines. Conversely, an explicit functional representation offers improved interpretability as well as better analytical tractability, which is imperative for downstream application. Hence, we utilize symbolic mathematics to obtain a closed-form formula for the activation function $\calA_2$ from the learned B-splines mapping.

   We consider a library of candidate functions containing polynomial, exponential, and trigonometric functions to approximate these B-splines.  For all batteries, we observe similar forms for the approximated representation with different values for coefficients $a,b,c,p,q,r,u,v$, and $w$ as:
    \begin{align}
        &\text{Polynomial:} \quad \quad \,\, \calA_2 \approx a - b(1-ck)^n, \\
        &\text{Exponential:} \qquad  \calA_2 \approx p - qe^{-rk}, \\
         &\text{Trigonometric:} \quad \calA_2 \approx u - sin/cos(v+wk).
    \end{align}
    Here, we consider 2$^{nd}$ to 5$^{th}$-order polynomials, i.e., $n \in [2,3,4,5]$ based on the lowest spline-order for our KAN models. Next, we evaluated these functional formulas for the learned activation function $\calA_{2}(k)$ in the range $k\in [0, \cdots, E]$ and estimated the corresponding $\soh$ using \eqref{eqn:dirA}. Table~\ref{tab:all_bat soh result} summarizes the performance of these 6 functional representation-based $\soh$ estimation in terms of MAE, MaxAE, root-mean-squared error (RMSE), and EOL error, along with the $R^2$ score for the functional approximation. $R^2$ score is a statistical measure that reflects the proportion of variability of the dependent variable captured by the model, and a higher value of $R^2$ score implies effective regression.
    
    \begin{figure}[b!]
    \centering
    \includegraphics[width=0.9\linewidth]{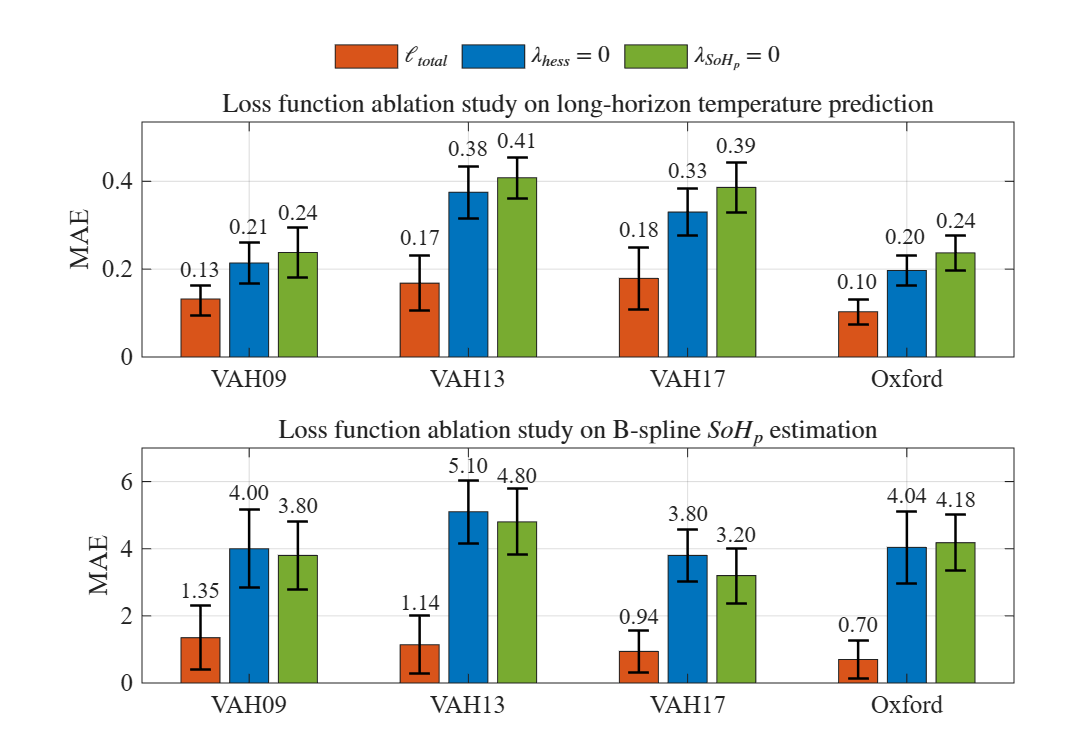}
    \caption{Plot compares the effect of regularization terms $\ell_{hess}$ and $\ell_{\soh}$ on temperature prediction (top plot)  and $\soh$ estimation (bottom plot) for  KAN models. }
    \label{fig:ablation}
\end{figure}

    Across all batteries, we observe slight increases in the error metrics for all five closed-form formulas compared to the learned B-spline representation,  as shown in Table~\ref{tab:all_bat soh result}. This additional error arises from the approximation of the  B-splines to explicit functional forms, which leads to a trade-off between estimation accuracy and analytical tractability of the $\calA_2$-based $\soh$ representation. Additionally, Table~\ref{tab:all_bat soh result} indicates that the cubic approximation yielded the lowest error values with the highest $R^2$ scores among the six functional forms for VAH09, VAH17, and Oxford batteries. For the VAH13 battery, the trigonometric $\soh$  estimation achieved the highest accuracy as well as the highest $R^2$ score. 
    We also approximated the $\calA_1$ function using the automatic symbolic fitting function, and observed a linear function of $\calA_1 \approx m \overline{T}(i) + s$ form for all batteries. This format matches the expected $\calA_1$ representation from battery physics as described in \eqref{eqn:discrete} and \eqref{eqn:A1}. Here,  the bias term $s$ maps to the ambient temperature effect $\xi \sum_{j=0}^{N-1} \gamma^j$.  This observation further illustrates the effective learning of the interpretable and physics-compliant 1-layer KAN models.

\subsection{Ablation study on loss function}
An ablation study on the tailored loss function $\ell_{total}$ can illustrate the impact of the proposed regularization terms $\ell_{hess}$ and $\ell_{\soh}$. Therefore, we retrain each KAN model with the hyperparameter configuration listed in Table~\ref{tab:training parameters}, while setting $\lambda_{hess} = 0$ first, and then $\lambda_{\soh} = 0$. We evaluate the performance of these models for each battery, and Fig~\ref{fig:ablation} captures the results of this ablation study.  The top and bottom plots, respectively, demonstrate the MAEs for temperature prediction and $\soh$ estimation under $\ell_{total}$, $\lambda_{hess} = 0$, and $\lambda_{\soh} = 0$.

Fig~\ref{fig:ablation} indicates that both $\ell_{hess}$ and $\ell_{\soh}$ regularization significantly affect the model performance. In particular, setting $\lambda_{\soh} = 0$ yields greater depletion in accuracy with higher MAE during temperature prediction. This result suggests that penalizing this regularization term ensures robust learning of the progressive heat generation with battery aging, resulting in improved predictive accuracy of the model. On the other hand, for VAH09, VAH13, and VAH17 batteries, $\ell_{hess}$ exhibits marginally larger effect on $\soh$ estimation, compared to the $\ell_{\soh}$ regularization. Thus, enforcing linear learning of $\calA_1$ essentially assists in reliable isolation of the degradation-driven increase in heat generation through $\calA_2$. Moreover, $\soh$ estimation is more sensitive to the removal of $\ell_{hess}$ and $\ell_{\soh}$ regularization, and exhibits a larger increase in MAEs compared to temperature prediction (shown in Fig~\ref{fig:ablation}). 

\subsection{Comparison with previous works}
To further validate the performance of the proposed SOH estimation method, we compared it with state-of-the-art thermal data-based SOH estimation methods \cite{tian2020state,ahuja2026lithium}. In \cite{ahuja2026lithium},  an empirical thermal model-based $\soh$ estimation algorithm is proposed that only uses surface temperature data and does not require internal battery parameters. They implemented their proposed algorithm using two observers, namely  Linear Quadratic Estimator (LQE) and Moving  Horizon Estimator (MHE), and validated it over the VAH09, VAH13, and VAH17 datasets. Our method outperformed both LQE and MHE  for the VAH09 and VAH13 data. For the VAH09 data, our best estimation RMSE is $1.65\%$ compared to $3.1\%$ with LQE and $2.0\%$ with MHE.   Similarly, for VAH13 data, our method achieved an RMSE of $1.76\%$, while LQE- and MHE-based $\soh$ estimations exhibited RMSEs of $2.0\%$ and $4.4\%$, respectively. Even our worst estimation errors for both VAH09 and VAH13 are smaller than both the LQE- and MHE-based estimation errors.
However,  for VAH17 data, MHE achieved a marginally higher accuracy, with $1.0\%$ RMSE compared to $1.11\%$ RMSE with our method. 

 We compare the performance of our method for the Oxford battery with \cite{tian2020state}. \cite{tian2020state} adopted support vector regression for $\soh$ estimation using the surface temperature measurement and reported an RMSE of $1.08\%$. Our best-performing B-spline $\soh$ estimation outperformed \cite{tian2020state} with a lower RMSE of $0.91\%$. However, we observed slightly larger RMSE with square and trigonometric $\soh$ estimation compared to \cite{tian2020state}. Overall, our proposed method exhibits improved or comparable performance with state-of-the-art thermal data-based SOH estimation methods. Additionally, our method has the clear advantage of identifying explicit functional relationships that can be leveraged for long-term prognostics or battery management. Unlike our method,  the current state-of-the-art data-driven methods \cite{tian2020state,ahuja2026lithium} solely provide SoH estimation using temperature data.

\section{Conclusion \& Future Work} \label{conc}
In this work, we propose a KAN-based data-driven pipeline to obtain a closed-form analytical representation of SoH degradation with the charging cycle number using battery temperature data.  We show that a simple 1-layer KAN structure can be utilized to learn and analytically represent the mapping from charging cycle number to battery heat generation, while only using the lumped battery temperature data. This heat generation expression can be utilized to obtain a closed-form analytical representation of  SoH degradation as a function of the charging cycle number. We also proposed a tailored loss function that penalizes prediction error as well as activation functions to ensure effective and physics-guided learning of the 1-layer KAN model, and our ablation study illustrates that this tailored loss function largely improves the predictive accuracy of the proposed method.  We evaluate the performance of our proposed method using two real-world datasets, each containing different battery chemistries and cycling conditions, including constant and dynamic current profiles at varied temperatures. We learn a dedicated KAN model for each battery, and in each scenario, the B-spline $\soh$ estimation from the KAN model closely follows the baseline $\soh$ calculated using current and voltage measurements. We also establish six closed-form representations of $\soh$ degradation by approximating the learned B-splines to polynomials, exponential, and trigonometric functions of the cycle number. Such explicit function-based $\soh$ estimation provides enhanced interpretability and analytical tractability with only a slight increase in error due to  B-spline approximation.

The proposed method thus presents a novel and explainable analytical representation of SoH that can be further explored and refined to capture the influence of other battery degradation factors. Since cycle number alone is often an incomplete aging indicator, to generalize the applicability of our proposed method, we will expand and fine-tune this baseline predictor to produce a robust  SOH estimation model, based on external stressors. In the future, we want to explore the transferability of the proposed approach across varied cell chemistries, cycling conditions, and ambient temperatures. We will also study the scalability of our KAN-based pipeline to battery packs to ensure practical application in electric vehicles or grid-scale energy storage systems.

\bibliography{ref1}

@article{ahuja2026lithium,
  title={Lithium-ion battery State of Health estimation based solely on temperature sensing},
  author={Ahuja, Nitisha and Moon, Jihoon and Bhaskar, Kiran and Rahn, Christopher D},
  journal={Journal of Power Sources},
  volume={666},
  pages={239068},
  year={2026},
  publisher={Elsevier}
}

@article{sun2024novel,
  title={A novel SOH estimation method with attentional feature fusion considering differential temperature features for lithium-ion batteries},
  author={Sun, Yiwen and Xie, Hengwei and Diao, Qi and Xu, Hongzhang and Tan, Xiaojun and Fan, Yuqian and Wei, Liangliang},
  journal={IEEE Transactions on Instrumentation and Measurement},
  volume={73},
  pages={1--11},
  year={2024},
  publisher={IEEE}
}

@article{gao2024soh,
  title={A SOH estimation method of lithium-ion batteries based on partial charging data},
  author={Gao, Renjing and Zhang, Yunfei and Lyu, Zhiqiang},
  journal={Journal of Energy Storage},
  volume={103},
  pages={114309},
  year={2024},
  publisher={Elsevier}
}

@article{zhang2023battery,
  title={Battery SOH estimation method based on gradual decreasing current, double correlation analysis and GRU},
  author={Zhang, Chaolong and Luo, Laijin and Yang, Zhong and Zhao, Shaishai and He, Yigang and Wang, Xiao and Wang, Hongxia},
  journal={Green Energy and Intelligent Transportation},
  volume={2},
  number={5},
  pages={100108},
  year={2023},
  publisher={Elsevier}
}

@article{zhou2021state,
  title={State-of-health estimation for LiFePO 4 battery system on real-world electric vehicles considering aging stage},
  author={Zhou, Litao and Zhao, Yang and Li, Da and Wang, Zhenpo},
  journal={IEEE Transactions on Transportation Electrification},
  volume={8},
  number={2},
  pages={1724--1733},
  year={2021},
  publisher={IEEE}
}

@article{gong2022state,
  title={State-of-health estimation of lithium-ion batteries based on improved long short-term memory algorithm},
  author={Gong, Yadong and Zhang, Xiaoyong and Gao, Dianzhu and Li, Heng and Yan, Lisen and Peng, Jun and Huang, Zhiwu},
  journal={Journal of Energy Storage},
  volume={53},
  pages={105046},
  year={2022},
  publisher={Elsevier}
}

@article{zhang2025cmmog,
  title={A CMMOG-based lithium-battery SOH estimation method using multi-task learning framework},
  author={Zhang, Chaolong and Tu, Liang and Yang, Zhong and Du, Bolun and Zhou, Ziheng and Wu, Ji and Chen, Liping},
  journal={Journal of Energy Storage},
  volume={107},
  pages={114884},
  year={2025},
  publisher={Elsevier}
}

@article{lin2012online,
  title={Online parameterization of lumped thermal dynamics in cylindrical lithium ion batteries for core temperature estimation and health monitoring},
  author={Lin, Xinfan and Perez, Hector E and Siegel, Jason B and Stefanopoulou, Anna G and Li, Yonghua and Anderson, R Dyche and Ding, Yi and Castanier, Matthew P},
  journal={IEEE Transactions on Control Systems Technology},
  volume={21},
  number={5},
  pages={1745--1755},
  year={2012},
  publisher={IEEE}
}

@article{yang2022lithium,
  title={Lithium-ion battery capacity estimation based on battery surface temperature change under constant-current charge scenario},
  author={Yang, Jufeng and Cai, Yingfeng and Mi, Chris},
  journal={Energy},
  volume={241},
  pages={122879},
  year={2022},
  publisher={Elsevier}
}

@article{tian2020state,
  title={State-of-health estimation based on differential temperature for lithium ion batteries},
  author={Tian, Jinpeng and Xiong, Rui and Shen, Weixiang},
  journal={IEEE Transactions on Power Electronics},
  volume={35},
  number={10},
  pages={10363--10373},
  year={2020},
  publisher={IEEE}
}

@article{zhang2026kan,
  title={What KAN mortality say: smooth and interpretable mortality modeling using Kolmogorov- Arnold networks},
  author={Zhang, Lianzeng and Zhuang, Yuan},
  journal={ASTIN Bulletin: The Journal of the IAA},
  pages={1--28},
  year={2026},
  publisher={Cambridge University Press}
}

@article{liu2024kan,
  title={\capitalisewords{Kan: Kolmogorov-arnold networks}},
  author={Liu, Ziming and Wang, Yixuan and Vaidya, Sachin and Ruehle, Fabian and Halverson, James and Solja{\v{c}}i{\'c}, Marin and Hou, Thomas Y and Tegmark, Max},
  journal={arXiv preprint arXiv:2404.19756},
  year={2024}
}

@article{sulaiman2024battery,
  title={\capitalisewords{Battery state of charge estimation for electric vehicle using Kolmogorov-Arnold networks}},
  author={Sulaiman, Mohd Herwan and Mustaffa, Zuriani and Mohamed, Amir Izzani and Samsudin, Ahmad Salihin and Rashid, Muhammad Ikram Mohd},
  journal={Energy},
  volume={311},
  pages={133417},
  year={2024},
  publisher={Elsevier}
}

@article{mallick2025kan,
  title={KAN-Therm: A lightweight battery thermal model using Kolmogorov-Arnold Network},
  author={Mallick, Soumyoraj and Ahamed, Faysal and Ghosh, Sanchita and Roy, Tanushree},
  journal={arXiv preprint arXiv:2509.09145},
  year={2025}
}

@article{ghosh2026kankoopman,
  title={KAN-Koopman Based Rapid Detection Of Battery Thermal Anomalies With Diagnostics Guarantees},
  author={Ghosh, Sanchita and Roy, Tanushree},
  journal={arXiv preprint arXiv:2602.21155},
  year={2026}
}

@article{mattia2025lithium,
  title={Lithium-ion battery thermal modelling and characterisation: a comprehensive review},
  author={Mattia, Luigi and Beiranvand, Hamzeh and Zamboni, Walter and Liserre, Marco},
  journal={Journal of Energy Storage},
  volume={129},
  pages={117114},
  year={2025},
  publisher={Elsevier}
}

@article{bills2023battery,
  title={A battery dataset for electric vertical takeoff and landing aircraft},
  author={Bills, Alexander and Sripad, Shashank and Fredericks, Leif and Guttenberg, Matthew and Charles, Devin and Frank, Evan and Viswanathan, Venkatasubramanian},
  journal={Scientific Data},
  volume={10},
  number={1},
  pages={344},
  year={2023},
  publisher={Nature Publishing Group UK London}
}

@article{preger2026capacity,
  title={Are Capacity and Energy Loss Equivalent Metrics for Battery Aging Reporting?},
  author={Preger, Yuliya and Wittman, Reed and Harris, Stephen J and Dubarry, Matthieu},
  journal={Journal of Electrochemical Energy Conversion and Storage},
  volume={23},
  number={2},
  pages={021109},
  year={2026},
  publisher={American Society of Mechanical Engineers}
}

@article{cheng2025soh,
  title={A SOH estimation method for lithium-ion batteries based on CPA and CNN-KAN},
  author={Cheng, Kaixin and Zhang, Chaolong and Shao, Kui and Tong, Jin and Wang, Anxiang and Zhou, Yujie and Zhang, Zhao and Zhang, Yan},
  journal={Batteries},
  volume={11},
  number={7},
  pages={238},
  year={2025},
  publisher={MDPI}
}

@article{zhang2024lithium,
  title={Lithium-ion battery SOH estimation method based on multi-feature and CNN-KAN},
  author={Zhang, Zhao and Liu, Xin and Zhang, Runrun and Liu, Xu Ming and Chen, Shi and Sun, Zhexuan and Jiang, Heng},
  journal={Frontiers in Energy Research},
  volume={12},
  pages={1494473},
  year={2024},
  publisher={Frontiers Media SA}
}

@article{christen2023theory,
  title={Theory of Fokker--Planck equations for reliability and remaining useful lifetime prognostics},
  author={Christen, Thomas and Macedo, Felipe},
  journal={IEEE Transactions on Reliability},
  volume={73},
  number={1},
  pages={344--356},
  year={2023},
  publisher={IEEE}
}

@article{chen2025estimation,
  title={Estimation of Lithium-Ion Battery SOH Based on a Hybrid Transformer--KAN Model},
  author={Chen, Zaojun and Lu, Jingjing and Wei, Qi and Wen, Jiayan and Wang, Yuewu and Li, Kene and Xu, Ao},
  journal={Electronics},
  volume={14},
  number={24},
  pages={4859},
  year={2025},
  publisher={MDPI}
}

@article{he2025soh,
  title={SOH Estimation Method for Lithium-Ion Batteries Using Partial Discharge Curves Based on CGKAN},
  author={He, Shengfeng and Qin, Wenhu and Yun, Zhonghua and Wu, Chao and Sun, Chongbin},
  journal={Batteries},
  volume={11},
  number={5},
  pages={167},
  year={2025},
  publisher={MDPI}
}

@inproceedings{liu2025soh,
  title={SOH Prediction of Battery Packs Using Dynamic Graph Convolution Combined with KAN-Driven Methods},
  author={Liu, Han and Wang, Yiming and Wang, Jiaju and Zhang, Ao and Yang, Haiqiang},
  booktitle={2025 4th International Conference on Green Energy and Power Systems (ICGEPS)},
  pages={129--134},
  year={2025},
  organization={IEEE}
}

@article{jarraya2025soh,
  title={SOH-KLSTM: A hybrid Kolmogorov-Arnold Network and LSTM model for enhanced Lithium-ion battery Health Monitoring},
  author={Jarraya, Imen and Atitallah, Safa Ben and Alahmed, Fatimah and Abdelkader, Mohamed and Driss, Maha and Abdelhadi, Fatma and Koubaa, Anis},
  journal={Journal of Energy Storage},
  volume={122},
  pages={116541},
  year={2025},
  publisher={Elsevier}
}

@article{jia2020soh,
  title={SOH and RUL prediction of lithium-ion batteries based on Gaussian process regression with indirect health indicators},
  author={Jia, Jianfang and Liang, Jianyu and Shi, Yuanhao and Wen, Jie and Pang, Xiaoqiong and Zeng, Jianchao},
  journal={Energies},
  volume={13},
  number={2},
  pages={375},
  year={2020},
  publisher={MDPI}
}

@inproceedings{yang2022prediction,
  title={Prediction Method of Remaining Service Life of Li-ion Batteries Based on XGBoost and LightGBM},
  author={Yang, Shipu},
  booktitle={2022 2nd International Conference on Algorithms, High Performance Computing and Artificial Intelligence (AHPCAI)},
  pages={324--327},
  year={2022},
  organization={IEEE}
}

@article{xiao2024state,
  title={State of health estimation for lithium-ion batteries using an explainable XGBoost model with parameter optimization},
  author={Xiao, Zhenghao and Jiang, Bo and Zhu, Jiangong and Wei, Xuezhe and Dai, Haifeng},
  journal={Batteries},
  volume={10},
  number={11},
  pages={394},
  year={2024},
  publisher={MDPI}
}

@article{chen2025residual,
  title={Residual useful life prediction of lithium-ion battery based on accuracy SoH Estimation},
  author={Chen, Guangwei},
  journal={Scientific Reports},
  volume={15},
  number={1},
  pages={6010},
  year={2025},
  publisher={Nature Publishing Group UK London}
}

@article{zhou2018prognostics,
  title={Prognostics for State of Health of Lithium-Ion Batteries Based on Gaussian Process Regression},
  author={Zhou, Di and Yin, Hongtao and Fu, Ping and Song, Xianhua and Lu, Wenbin and Yuan, Lili and Fu, Zuoxian},
  journal={Mathematical Problems in Engineering},
  volume={2018},
  number={1},
  pages={8358025},
  year={2018},
  publisher={Wiley Online Library}
}

@article{berecibar2016critical,
  title={Critical review of state of health estimation methods of Li-ion batteries for real applications},
  author={Berecibar, Maitane and Gandiaga, I{\~n}igo and Villarreal, Irune and Omar, Noshin and Van Mierlo, Joeri and Van den Bossche, Peter},
  journal={Renewable and Sustainable Energy Reviews},
  volume={56},
  pages={572--587},
  year={2016},
  publisher={Elsevier}
}

@article{sun2024safety,
  title={Safety behaviors and degradation mechanisms of aged batteries: A review},
  author={Sun, Shuguo and Xu, Jun},
  journal={Energy Materials and Devices},
  volume={2},
  number={4},
  pages={9370048},
  year={2024},
  publisher={清华大学出版社}
}

@article{paradiso2026failure,
  title={Failure mechanisms and risk assessment methodologies for lithium-ion batteries: A comprehensive review},
  author={Paradiso, Silvia and Tulabi, Milad and Bubbico, Roberto},
  journal={Journal of Energy Storage},
  volume={162},
  pages={122130},
  year={2026},
  publisher={Elsevier}
}

@article{howey2017oxford,
  title={Oxford battery degradation dataset 1},
  author={Howey, D and Birkl, C},
  year={2017},
  publisher={University of Oxford}
}

@article{andre2006real,
  title={Real-world European driving cycles, for measuring pollutant emissions from high-and low-powered cars},
  author={Andr{\'e}, Michel and Joumard, Robert and Vidon, Robert and Tassel, Patrick and Perret, Pascal},
  journal={Atmospheric Environment},
  volume={40},
  number={31},
  pages={5944--5953},
  year={2006},
  publisher={Elsevier}
}

@article{li2024soh,
  title={SOH estimation method for lithium-ion batteries based on an improved equivalent circuit model via electrochemical impedance spectroscopy},
  author={Li, Chaofan and Yang, Lin and Li, Qiang and Zhang, Qisong and Zhou, Zhengyi and Meng, Yizhen and Zhao, Xiaowei and Wang, Lin and Zhang, Shumei and Li, Yang and others},
  journal={Journal of Energy Storage},
  volume={86},
  pages={111167},
  year={2024},
  publisher={Elsevier}
}

\end{document}